\documentclass[onecolumn]{article}
\usepackage{amsmath,amsfonts}
\usepackage{algorithmic}
\usepackage{amsmath,amssymb,amsfonts}
\usepackage{multirow} 
\usepackage{array}
\usepackage{stfloats}
\usepackage{url}
\usepackage{graphicx}
\usepackage{caption}
\usepackage[margin=3cm]{geometry}
\usepackage[dvipsnames]{xcolor}

\begin{document}

\title{Design of a Microprocessors and Microcontrollers Laboratory Course Addressing Complex Engineering Problems and Activities}

\author{Fahim Hafiz$^1$, Md. Jahidul Hoq Emon$^1$, Abid Hossain$^1$,\\ Md. Saddam Hossain Mukta$^{2}$, Salekul Islam$^{1,3}$ and Swakkhar Shatabda$^{4}$,\\ $^1$Department of Computer Science and Engineering, United International University\\ Plot-2, United City, Madani Avenue, Badda, Dhaka-1212\\ $^2$LUT School of Engineering Sciences, Lappeenranta-Lahti University of Technology\\ Yliopistonkatu 34, 53850 Lappeenranta, Finland\\ $^3$Department of Electrical and Computer Engineering, North South University\\ Plot \# 15 Block B, Bashundhara R/A, Dhaka‐1229, Bangladesh\\ $^4$Department  Computer Science and Engineering, BRAC University\\ Kha 224, Bir Uttam Rafiqul Islam Ave, Dhaka 1212, Bangladesh\\ \footnote{Email: \{fahimhafiz,jahidul,abid\}@cse.uiu.ac.bd, saddam.mukta@lut.fi,salekul.islam@northsouth.edu, swakkhar.shatabda@bracu.ac.bd}
}

\date{}
% \markboth{IEEE Transactions on Education}%
% {}


%\history{Date of publication xxxx 00, 0000, date of current version xxxx 00, 0000.}
%\doi{10.1109/ACCESS.2017.DOI}

\maketitle

\begin{abstract}

\textit{Background:}
Microprocessors and microcontrollers are ubiquitous in modern technology, driving applications across diverse fields. To prepare future engineers for Industry 4.0, effective educational approaches are crucial in this technology.

\textit{Purpose:}
This paper proposes a novel curriculum for the microprocessors and microcontrollers laboratory course. The proposed curriculum blends structured laboratory experiments with an open-ended project phase while addressing complex engineering problems and activities.

\textit{Design:}
Traditional microprocessors and microcontrollers curricula often fail to capture the complexity of real-world applications encountered in Industry 4.0. This curriculum addresses this gap by incorporating experts' insights from both industry and academia. It trains students with the necessary skills and knowledge to thrive in this evolving technological landscape, preparing them for success upon graduation. After designing the course, we have taken feedback from the students which showed positive learning curve and experiences regarding this designed curriculum.

\textit{Results:}
The proposed lab enables students to perform hands-on experiments using advanced microprocessors and microcontrollers where they can leverage their acquired knowledge by working in teams to tackle self-defined complex engineering problems that utilize these devices and sensors. Statistical analysis shows the proposed curriculum significantly improves student learning outcomes, particularly in their ability to both formulate and solve complex engineering problems as well as engaging in complex engineering activities.

\textit{Conclusions:}
This curriculum integrates project-based learning, where students define complex engineering problems for themselves. Additionally, this proposed method fosters multidisciplinary learning and equips students with problem-solving skills that can be applied in real-world scenarios. This approach actively engages students, fostering deeper understanding and enhancing their learning capabilities.

% Microprocessors and microcontrollers are ubiquitous in modern technology, driving applications across diverse fields. To prepare future engineers for Industry 4.0, effective educational approaches are crucial. This paper proposes a novel curriculum for a microprocessors and microcontrollers laboratory course. The proposed curriculum blends close-ended lab experiments with an open-ended project phase. Students gain foundational knowledge through experiments using advanced microprocessors and microcontrollers. Subsequently, they leverage this knowledge in teams to tackle a self-defined complex engineering problem using these devices and sensors. This approach fosters multidisciplinary learning and equips students with problem-solving skills applicable to real-world scenarios. The curriculum design incorporates insights from industry-academic experts and addresses the need for effective education in the context of Industry 4.0.
\end{abstract}


\maketitle

\section{Introduction}
\label{sec:introduction}

Microprocessors and microcontrollers offer high computing power with less cost \cite{barrett2022microcontrollers} and have a broad range of applications from industrial fields \cite{aliu2012design}, automotive \cite{van2021renesas, biba2020power} fields to medical fields \cite{nimunkar2020microcontrollers, ayemtsa2024medical, diab2022embedded}. IoT-based projects are typically designed using such microcontrollers and have vast applications in home appliances too \cite{ahsan2021smart, sisavath2021design}. Due to the various applications, the total global microcontrollers' market will reach 117 billion approximately by the end of 2025 \cite{EpoSS}, and will continue to grow due to the continuous integration of machine learning (ML) and computer vision with such micro-devices \cite{9893137}. Microprocessors such as Raspberry Pi (RPi) are equipped with RAM and external memory. As a result, such microprocessors are very small in size but work just like a high-computing computer. In recent years, lite ML models such as TensorFlow-lite and tiny ML \cite{warden2019tinyml, besimi2020using} allow seamless computing of high-end neural network models on microprocessors and microcontrollers such as Arduino, ESP-32, ARM, RPi, etc. However, within the context of Industry 4.0, there is a demand for pioneering technologies capable of seamlessly integrating industry, humanity, and all aspects of daily life to serve a greater purpose. However, the transition to Industry 4.0 will take time and to accelerate this phase, development in the engineering education sector should take place to produce high-quality engineers \cite{raikar2018upsurge}. From that perspective, engineering students should experience a multidisciplinary approach to learning different hardware platforms such as Arduino, RPi, etc.\ while integrating software \cite{7344259}.

There have been various strategies and approaches in the present education system to teach undergraduate students different microprocessors and microcontrollers. This lab course is usually mandatory for electrical and computer science engineering students all over the world.  Initially, it was mainly taught from the theoretical perspective \cite{1}. However, this course is presently taught from the application perspective \cite{1, 2}. Recently, this course has been taught in a multidisciplinary approach to engineering students. Students from electrical engineering, computer science, biomedical engineering or mechanical engineering can be taught this course due to its multidisciplinary nature \cite{3}. Such a multidisciplinary approach can be very effective to learn industry-level applications and students can be motivated through the level of outcomes they will attain in this approach. Project-based learning approaches in embedded systems and microcontrollers are also taken in various universities \cite{4, 5}. In this approach, students are provided with all the information of the project and learn the different aspects of the course while doing such practical projects \cite{4}. This project-based approach showed effectiveness in learning the details of microprocessors and microcontrollers rather than designing new applications using them \cite{4, 5}. In recent years, we have seen different microprocessors and microcontrollers being taught in this course such as Arduino, AVR, RPi, ARM, ESP32, etc.\ \cite{6, 7, 19, 20, 24}. All of these processors have different functionalities and can be used in various aspects depending on the complexity of the engineering problems. Regardless of the types of microprocessors and microcontrollers, importance should be given to how the course is designed to effectively teach these hardware platforms to students of diverse backgrounds. 

After doing an extensive literature review, we conclude two types of design while teaching such multidisciplinary courses: i) close-ended approach, and ii) open-ended approach. The close-ended approach includes basic lab experiments where students understand the hardware prototypes and perform different group-wise tasks after learning the theoretical and applied knowledge of this hardware from the instructor. In this approach, the students are usually evaluated from different lab tasks, written and practical exams, lab reports, etc. On the other hand, the open-ended approach \cite{newman2003open} is where students are provided with the necessary materials and guidelines on the hardware prototypes. Based on this information, the students first select a real-life problem, solve the problem in a team, and learn the different aspects of the prototypes while completing the real-life project. Note that an open-ended project has no obvious solution, multiple solutions may exist and students have to come up \textcolor{black}{with} their own solutions. 

In this work, we have proposed and designed a new curriculum for the microprocessors and microcontrollers lab at a Computer Science and Engineering Department. The course code and title of this lab is {CSE 4326: Microprocessors and Microcontrollers Laboratory} which has already a previous course curriculum. This lab course is taught at the 8th-semester students and has {CSE 2124: Electronics Laboratory} as a prerequisite. \textcolor{black}{In the {CSE 2124: Electronics Laboratory} course, the students use Proteus simulation software to conduct the experiments in a virtual environment. Therefore, Proteus gives the students a significant advantage in simulating the experiments of the newly designed course.} 
 Our new lab course is mandatory for CSE students where they have to learn the basic properties of microprocessors, and microcontrollers, and how they can be interfaced with other hardware systems to solve real-life problems. We have proposed a mixture of close-ended (lab experiments) and open-ended (project-based) multidisciplinary approaches while teaching the students about some of the most advanced microprocessors and microcontrollers. In this proposed approach, several lab experiments are designed for the students to understand and gain in-depth knowledge of microprocessors, microcontrollers, and different hardware devices. Based on the knowledge gained from these close-ended lab experiments, the students are required to propose a complex engineering problem by forming a team to solve real-world problems using microprocessors, microcontrollers, and other sensors. The complex engineering problem proposed by the students should be well-researched before formulating the problem's definition. Note that we are following the definition of complex engineering problem (presented in Table~\ref{tab:cea}) given by the Washington Accord (the largest, international accreditation body for engineering programs). In this way, the students will understand how to solve an open problem based on the curriculum taught at the lab. This approach is open-ended approach according to our proposed course curriculum. Before designing this new curriculum, we performed extensive research and took industry-academic experts' opinions which are provided in the following sections.

The rest of the paper is organized as follows. Section \ref{s:lit-rev} summarizes the similar efforts found in the literature with a categorization. Section \ref{s:dev-cur} presents the previous curriculum with the drawbacks found and the feedback taken from different stakeholders on designing a new curriculum. Section \ref{s:propos-curcl} presents the proposed curriculum including the newly designed guided experiments and open-ended project. Section \ref{s:result-analys} summarizes the results including students' survey results and the performed statistical analysis and qualitative analysis of the open-ended project. Finally, section \ref{s:con} concludes the paper \textcolor{black}{by} mentioning some possible future works. 

\section{Literature Review} \label{s:lit-rev}
In this section, we review some of the relevant works \textcolor{black}{that are} similar to our work. A new course was designed and shown in the work \cite{19} at the University of Calgary, using RPi to make the learning mechanism of assembly language enjoyable and interactive for the students. In the RPi part of the course, students learn to create an interactive video game using ARM assembly language. The author showed that the distribution of grades in the exams improved significantly after introducing RPi to the curriculum. A voluntary, anonymous survey collected feedback from 198 students in two consecutive semesters, indicating an overall response rate of 58\%, evaluating aspects like enjoyment, learning experience, motivation, and independent learning in the new course format. Quantile-Quantile normality test was performed before and after introducing RPi which showed that the mean grades after introducing RPi were not normally distributed compared to the previous curriculum. Consequently, the authors performed a t-test which revealed that the integration of RPi was significant. Two surveys were also conducted concerning RPi and future studies. In both of them, the students expressed that RPi enhanced their learning curve and they would like to explore more about low-level and game programming in the future. In \cite{20}, the authors designed a MATLAB course while integrating RPi to enhance the learning capability of the BME students while handling real-life data. They proposed six experiments and performed both numerical course evaluation and survey on the students before and after incorporating RPi. After including RPi, the evaluation score increased. The student enjoyed programming more after introducing such hardware experiments and the authors indicated that as the BME students in the senior year have to do more complex real-life analysis, this newly proposed course would help them to understand how real-life projects work.

\begin{table*}[!htb]
\centering
    \caption{Categorization of the present literature review}
    \label{tab:literature}
    \begin{tabular}{|p{0.1\textwidth}|p{0.02\textwidth}|p{0.3\textwidth}|p{0.2\textwidth}|p{0.12\textwidth}|p{0.12\textwidth}|}
    \hline \textbf{Discussed Area/Study Type} & \textbf{Ref.} & \textbf{Learning Emphasis} & \textbf{Assessment Methods} & \textbf{Main Focus/Outputs} & \textbf{Published In}\\
    \hline %\multirow{5}{*}
    {CS/Engineering Courses with hardware} & \cite{19}& Development of Capabilities &Lab Evaluations, Assessments,Exams&Multidisciplinary & SIGCSE - ACM Digital Library\\ \cline{2-6}
    
    & \cite{20}& Real-life Learning Capability Enhancement for BME Students, data processing &Course Evaluation and Exams &Multidisciplinary & 2017 ASEE Annual Conference \& Exposition\\ \cline{2-6}

    & \cite{23}& Python-based skill development using RPi for enhance understandings &Lab evaluations and self-developed
    evaluation sheet  &Multidisciplinary & Heliyon\\ \cline{2-6}
    
    & \cite{24}& Project-based Computing Learning capabilities with integration of RPi &Project Evaluation, Student Evaluation
    & Multidisciplinary and Project-based & SIGCSE - ACM Digital Library\\ \cline{2-6}
    
    & \cite{ShekharUnpack} & Determining the Motivating factors of the individuals interested in Engineering and CS Undergraduate Levels &Project-based Learning and Teamwork & Multidisciplinary and Project-based & IEEE Transactions on Education\\ \cline{2-6}

    \hline %\multirow{3{*}
    {Course Curriculum Design} & \cite{21}& Practical Skills Enhancement in I\&C System with Real-life problem-solving capabilities according to the stakeholders &Specific Indicators based Evaluation such as Written Exam, Viva-Voce, Practical Examination, Project report, Survey & Multidisciplinary and Project-based & IEEE Transactions on Education\\ \cline{2-6}
    
    & \cite{22} & Designing real-life industrial Robotic Prototypes &Written Exams, Lab practicals, Students Participations,  and Lab reports &Multidisciplinary& IEEE Transactions on Education\\ \cline{2-6}

    & \cite{chiu2021creation} & AI Curriculum for the school students & Survey and Questionnaires & Multidisciplinary & IEEE Transactions on Education\\

    \hline %\multirow{2}{*}
    {Student Feedback Analysis on Existing Course} & \cite{ZhaoGame} & Game-based learning approach to teach programming in undergraduate levels &Demographic Questionnaires
    &Interactive with feedback mechanism& IEEE Transactions on Education\\ \cline{2-6}
    
    &  \cite{riese2023engineering} & Students Experience Analysis in CS Courses &Surveys considering distinct elements of engineering course &Multidisciplinary and Project-based & IEEE Transactions on Education \\

    \hline
    \end{tabular}
    
\end{table*}

In \cite{21}, the authors proposed a new curriculum for teaching the Instrumentation and Control Systems (I\&C System) for engineering students with four intended learning outcomes. During prototype planning, they contacted I\&C systems industry experts and asked questions relevant to the learning outcomes of their course. According to the experts' survey, the prototype was designed for the students. The design of such a prototype has a multidisciplinary approach keeping the feedback from the industrial organizations' survey. As per the authors’ knowledge, and from studies, such prototypes do not exist in other Indian institutes. Both course structure and criteria were proposed with respect to the ABET criteria and survey results of the industrial leaders. The students were divided into groups to learn and share the lab experiments in a collaborative environment. Two types of evaluations were performed after deploying the course: student survey and evaluation based on specific indicators. Students were asked 10 questions each having ratings from 1 to 5. The authors explained the poor rating in some of the questionnaires. The course has five specific indicators (SI) on which the evaluation was performed. Overall evaluation of this course was based on the Written Exam, Viva-Voce, Practical Examination, and Project report where these segments were mapped to different SIs. A detailed semester-wise comparison was performed based on these SI’s before and after introducing this prototype. The survey questionnaire for the students and their answers were categorized into different Specific Indicators. This was done for the students attending the newly designed course and the previous three years for the old courses. The average of the specific indicators increased in the semester of introducing this new prototype. Also, Cronbach’s $alpha$ test and \textit{t-test} were performed on these specific indicators which shows the reliability and effectiveness of this prototype respectively.  Another work \cite{22} designed a mechatronics course while maintaining a multidisciplinary approach as well as the ABET criterion for an undergraduate program at the Automation Department at Universidad Autónoma de Querétaro, Querétaro, Mexico. The authors focused on teaching and building a robot prototype (a robot arm) for the students. They have prepared six practical sessionals for teaching such industrial robot prototypes which included basic levels of familiarization of robot prototypes to advanced levels of controlling the robotic arm in wireless mode. Two types of evaluations were proposed after teaching the practical sessions which are Student Surveys and Evaluation Based on Specific Indicators (SIs). Individual and team wise (5 students per group) survey were performed for 10 questions each having 5 points (1 means strongly disagree, 5 means strongly agree). The survey showed on average 4 to 5 points while having low standard deviation for most of the questions. On the other hand, the course was designed keeping in mind 5 SIs, and the evaluation for the grading was divided into four parts: Written Exam (20\% marks), Practicals (50\% marks), Lab Report (20\% marks), Student’s participation and their ability to function in a multi-disciplinary team (qualitative-10\% marks) based on these SIs. This prototype was introduced in 2016 and the average number (percentage) in each specific indicator was calculated for three consecutive years: 2014, 2015, and 2016. It was found that the average percentage number seems to improve in the year 2016 compared to the previous prototype which shows that the introduced robot prototype is more effective for the students. However, the authors mentioned that taking surveys from the stakeholders matters too in such analysis which they did not include in the performance analysis of the new prototype.

Kawash et al.~\cite{23} proposed a Python-based lab course that has been designed to perform image and video processing. The course included seven experiments and used RPi 3 model B for a basic understanding of image and multidimensional data like interpolation, feature detection, segmentation, and transformation. The lab course has been evaluated by the participating students in the evaluation system of the university and a self-developed evaluation sheet. Overall, in both evaluation systems, the lab course has been rated between excellent and very good. Another work \cite{24} proposed a project-based lab course for first-year undergraduate students to learn computing using RPi. The RPi allowed students to perform projects that require a physical computing aspect by using the GPIO headers on the Pi. The proposed lab task included implementing algorithms, mapping and graphing the weather, Twitter posting, and scraping. Assignments in this lab were mostly study-based. They were given topics or videos to study and discuss during discussions. The project marking criteria included application-based project ideas related to the area of computing, relative complexity, and effort, project completion status according to the original proposal, presentation of Project as well as the individual reflection. Both project evaluation and student evaluation were performed which showed that the student found the technology tools utilized were appropriate to the course as well as received positive feedback. 

Other than course design, \textcolor{black}{another} study \cite{riese2023engineering}  focused on how engineering students (non-CSE majors) experience and perceive different aspects of the introductory CS courses. The authors surveyed the students taking six different CS courses where the survey questionnaires focused on 5 different attributes of the course: Lab assignments, Exams, Individual Projects, Course Coordinators, and Teaching Assistants. These attributes influence the learning curve and experience of the students. From these surveys, one of the vital findings was that the students perceived lab assignments as a learning activity rather than “lab assignment as a necessary evil” \cite{inproceedingsRiese}. On the other hand, the project in these courses was experienced as the most enjoyable and evocative one as they solved real-life problems in the projects \cite{riese2023engineering}. However, they also found a gap between the projects and lab experiments. That’s why collaborative teamwork seemed a viable option to them while working on such projects \cite{riese2023engineering}. Another study \cite{ShekharUnpack} was conducted in this paper to find the factors that can motivate high school students in engineering and computer science subjects leveraging project-based learning methods. Motivation in engineering education brings better outcomes from engineering courses. \textcolor{black}{Conversely}, several studies \cite{Millsarticle}, \cite{Chenarticle} suggest that project-based learning is more effective in engaging students to participate in teamwork and motivates them to bring out their best. The authors formulated six diversified focus groups from four different schools in different US Schools. First, the authors \cite{ShekharUnpack} proposed some technical classes on robotics, embedded systems, and IoT-based systems using Arduino, and RPi with Python programming for the focus groups. Students attended these classes and learned basic hardware-software interfacing which enables them to apply basic programming logic to hardware while conducting some real-life projects. Then a survey was conducted which suggested that, while complex projects challenged the students to think deeper and boosted their confidence, problem definitions in the project should be made clear for them. Multidisciplinary approaches such as cross-combining mathematics, programming, and other subjects were found effective in motivating the students. However, project-based learning became one of the key motivating factors according to the survey where the students worked in a team to solve real-life problems which were relevant contextually. Zhao et al. \cite{ZhaoGame} proposed a game-based approach to create an interactive environment with the students to teach programming knowledge to the students. Over 100 students from three institutions studying at first and second-year undergraduate levels participated in the pilot project of the proposed approach where the evaluation was quantified using demographic questionnaires. The findings indicated that the game-based approach helped students to learn the concepts in an interactive environment while captivating their concentration.  However, step-by-step instructions are needed in such a game-based approach, otherwise the user experience can be affected severely. Also, learning capabilities vary from person to person where personalization is needed in the proposed games. These types of game-based learning lack personalization whereas project-based learning motivates personalized learning capabilities. Chiu et al.~\cite{chiu2021creation} discussed creating a pre-tertiary AI curriculum for students in Hong Kong schools. The objective of creating such an AI curriculum for the students was to motivate them to pursue this multidisciplinary technology in the future. The curriculum framework consists of beginner levels to advanced levels while incorporating knowledge of AI, application-based tasks as well as the ethical side of AI, overall delivering the future impact of AI towards these young minds. The authors developed a Robot car, "CUHKiCar" which allows the students to perform AI experiments. Also, reprogramming options allow the students to learn different functionalities of the AI which enables interactive learning. The proposed curriculum in this work is flexible for the teachers and students offering versatile learning approaches. The curriculum was also evaluated after surveys which showed that students became motivated to learn AI, and solve complex projects after completing this curriculum.    
Other relevant works \cite{25, 26, 27, 28} proposed different engineering lab courses while maintaining a similar strategy as discussed in previous literature above.

The majority of previous methods have taken multidisciplinary approaches while designing the courses. However, they have usually taken close-ended approaches which indicate teaching different experiments and evaluating the students based on these experiments. On the other hand, some studies proposed open-ended (project-based) approaches. The overall scenario of our literature review is divided into several segments shown in Table~\ref{tab:literature} where we can see the main focus or outputs of the different studies are multidisciplinary and project-based approaches. However, the assessments in the course design also included close-ended evaluations.

In our work, we have taken a mixture of \textit{close-ended} and \textit{open-ended} approaches for more effective learning. In this approach, we designed the experiments while keeping the multidisciplinary perspective as well as the feedback from the industry and academic experts.

\section{Method} \label{s:dev-cur}

The new curriculum has been developed by taking feedback from industry and academic experts. Here, we present the limitations of the previous curriculum and then explain how we have designed the new one considering that feedback. A general framework of our overall approach while designing the new curriculum is illustrated in Fig. 1. We first analyzed the overall quality of the present curriculum of this course and determined the drawbacks at the faculty level. Then, stakeholder participation was initiated to identify the industry demand in the field of microprocessors, microcontrollers, and relevant fields. We then address the overall findings and suggestions from the experts while proposing the curriculum. In the later part of this paper, we will discuss this approach in detail.

\begin{figure}[h!]
    \centering
    \includegraphics[width=0.35\textwidth]{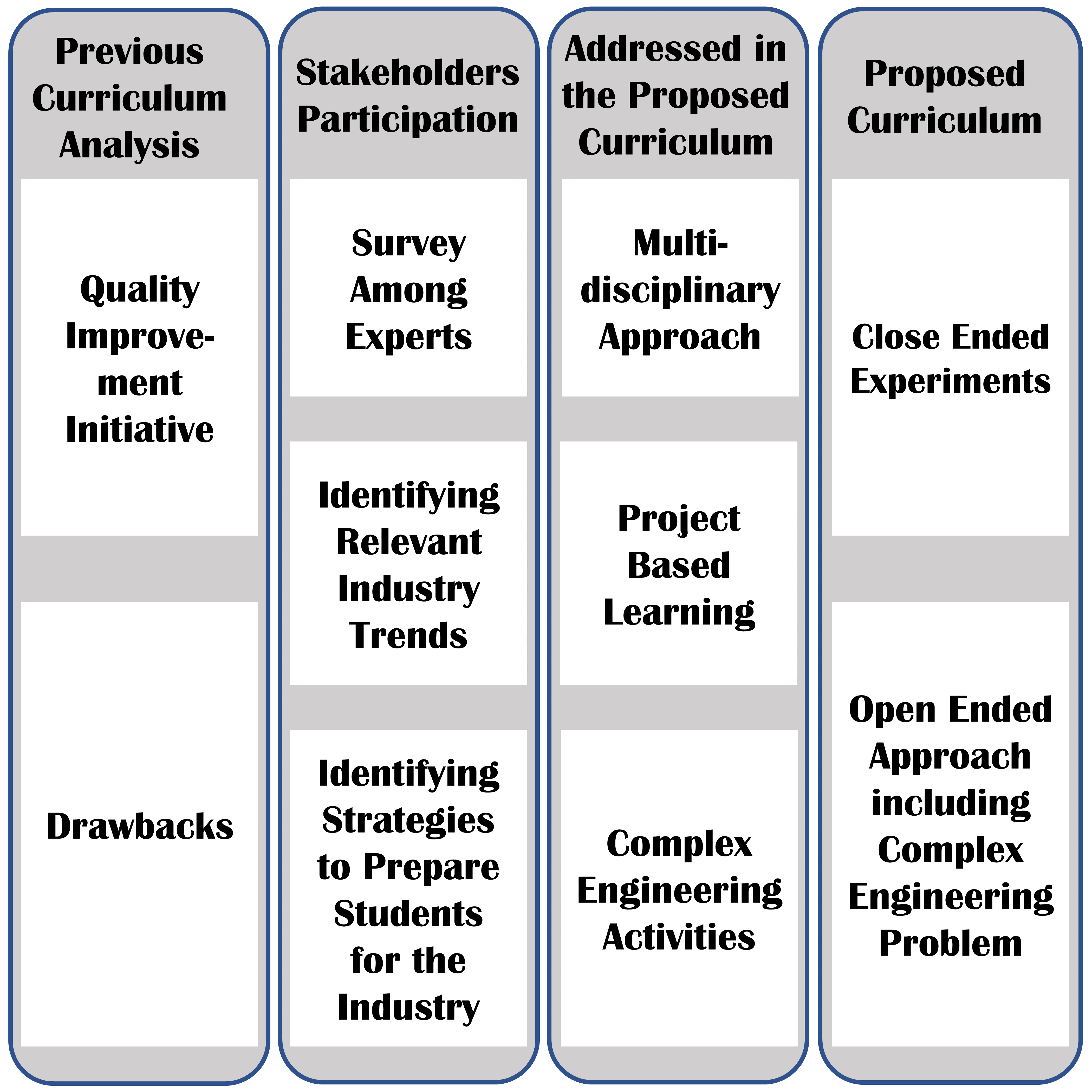}
    \caption{General Framework of our approach to design the new curriculum.}
%\caption{Example of a parametric plot}
    \label{fig:industry_microntler}
\end{figure}

\subsection{Previous curriculum and their drawbacks}
In our previous curriculum, the students were taught the AVR system and Arduino. There were seven experiments taught in the laboratory. A brief of the experiments are given in Table~\ref{tab:prev}.

\begin{table}[!htb]
\centering
    \caption{Short overview of the Experiments in Previous Curriculum}
    \label{tab:prev}
    \begin{tabular}{|p{0.06\textwidth}|p{0.25\textwidth}|p{0.4\textwidth}|}
    \hline \textbf{Exp. No.} & \hspace{1.3cm}\textbf{Experiment Title} & \hspace{2.5cm} \textbf{Objectives} \\
    \hline \vspace{0.5cm}  \hspace{0.5cm}1 & \vspace{0.35cm} Microcontroller Configuration and LED Interfacing with ATmega 32 &  To learn how to identify the I/O (input/output) pins and ports in the microcontroller, how to use an AVR programmer to write HEX code into the ATmega 32 microcontroller, and how to simulate a project in Proteus.\\
    \hline \vspace{0.5cm} \hspace{0.5cm} 2 & \vspace{0.1cm}Microcontroller Input Output Operation & To learn  how to configure a microcontroller pin as input and output, how to design the pull-up circuit, and how to solve the switch bounce problem.\\
    \hline \vspace{0.5cm} \hspace{0.5cm} 3 & \vspace{0.1cm}Seven Segment Display Interfacing with ATmega 32 &To learn how to explain the difference between cathode and anode seven-segment displays and how to interface multiple seven-segment displays with ATmega32. \\
    \hline \vspace{0.3cm} \hspace{0.5cm} 4 &\vspace{0.1cm} LCD Display Interfacing with ATmega 32 & To learn how to configure LCD display and how to interface LCD display with ATmega32 and show custom characters. \\
    \hline \vspace{0.5cm} \hspace{0.5cm} 5&\vspace{0.2cm} Introduction to Arduino Uno Board&To learn how to configure digital pins in an Arduino board, how to apply serial communication between the Arduino board and a computer for data transmission, and how to interface the push button with Arduino. \\
    \hline \vspace{0.5cm} \hspace{0.5cm} 6 &\vspace{0.2cm} Analog Input-Output using Arduino Uno Board&To learn how to take readings from analog sensors, how to use analog output (Pulse Width Modulation (PWM)) to fade in/out an LED, the function of Analog to Digital Converter (ADC) and Digital to Analog Converter (DAC). \\
    \hline \vspace{0.5cm} \hspace{0.5cm}7 & Receive Humidity and Temperature Update from One Arduino Board to Another Arduino Board using I2C Communication Protocol& To learn how to configure the I2C communication protocol in Arduino, how to interface the DHT11 sensor with the Arduino board, and how to apply the I2C communication protocol for transmitting and receiving data. \\
    
    \hline
    \end{tabular}
    
\end{table}

 The course outcomes of the previous curriculum with description are given in Table~\ref{tab:CO-PO} with the Course Outcomes and the Program Outcomes (POs) that are mapped to the course via course outcomes are given in Table~\ref{tab:CO-PO} and the PO statements are shown in Table~\ref{tab:POs}. 
 
\begin{table}[!htb]
\centering
    \caption{Course Outcomes (COs) with Description.}
    \label{tab:CO-PO}
    \begin{tabular}{|c|p{0.33\textwidth}|p{0.1\textwidth}|p{0.1\textwidth}|}
    \hline \textbf{COs} & \textbf{Description} & \textbf{Program Outcome} & \textbf{Bloom's Domain}\\
    \hline CO1 & Illustrate the 
interfacing of microcontroller with different input/output devices and simulate using Proteus to observe and analyze behaviors of different electronic hardware circuits &  e & C\\
    \hline  CO2 & Design a basic 
embedded hardware group project & c & A\\
    \hline CO3 & Work in a team and communicate effectively.
 &i,j \ &  A\\
    
    \hline
    \end{tabular}
    
\end{table}

\begin{table}[!htb]
    \centering
    \caption{Program Outcome (PO) Statements that are mapped to this course via course outcomes.}
    \label{tab:POs}
\begin{tabular}{|p{0.1\textwidth}|p{0.5\textwidth}|}
\hline
\bf \hspace{1cm}PO&\bf \hspace{2.5cm} Statement\\
\hline
     \hspace{.9cm}c&Design solutions for complex engineering problems and design systems, components, or processes that meet specified needs with appropriate consideration for public health and safety, cultural, societal, and environmental considerations. (K5)  \\
     \hline
     \hspace{.9cm}e&Create, select, and apply appropriate techniques, resources, and modern engineering and IT tools, including prediction and modelling, to complex engineering problems, with an understanding of the limitations. (K6)\\
     \hline
     \hspace{.9cm}i&Function effectively as an individual, and as a member or leader in diverse teams and in multi-disciplinary settings.\\
     \hline
     \hspace{.9cm}j&Communicate effectively on complex engineering activities with the engineering community and with society at large, such as being able to comprehend and write effective reports and design documentation, make effective presentations, and give and receive clear instructions.\\
     \hline
\end{tabular}

\end{table}

Note that CO1 and CO2 are mapped to POc and POe which require the ability of students to solve complex engineering problems and POj which requires communication on complex engineering activities. POc and POe are mapped to two knowledge profiles K5 and K6. These definitions are provided in Graduate Attributes and Professional Competencies defined by Washington Accord. For the sake of comprehension, the definitions of Knowledge Profile (KP), Complex Engineering Problem (CEP) solving, and Complex Engineering Activities (CEA) are given in Table~\ref{tab:kp}, Table~\ref{tab:cep} and Table~\ref{tab:cea} respectively.

\begin{table}[!htb]
    \centering
    \caption{Knowledge Profile}
   \begin{tabular}{|p{0.1\textwidth}|p{0.5\textwidth}|}
    \hline
    & \bf Attribute\\
    \hline
    K1&A systematic, theory-based understanding of the natural sciences applicable to the discipline\\
    \hline
    K2&Conceptually based mathematics, numerical analysis, statistics and the formal aspects of computer and information science to support analysis and modeling applicable to the discipline\\
    \hline
    K3&A systematic, theory-based formulation of engineering fundamentals required in the engineering discipline\\
    \hline
    K4&Engineering specialist knowledge that provides theoretical frameworks and bodies of knowledge for the accepted practice areas in the engineering discipline; much is at the forefront of the discipline\\
    \hline
    K5&Knowledge that supports engineering design in a practice area\\
    \hline
    K6& Knowledge of engineering practice (technology) in the practice areas in the
engineering discipline\\
    \hline
    K7& Comprehension of the role of engineering in society and identified issues in
engineering practice in the discipline: ethics and the engineer’s professional responsibility to public safety; the impacts of engineering activity; economic, social, cultural, environmental and sustainability\\
    \hline
    K8&Engagement with selected knowledge in the research literature of the discipline
      \\
    \hline
    \end{tabular}
    \label{tab:kp}
\end{table}

\begin{table}[!htb]
    \centering
    \caption{Range of Complex Engineering Problem Solving}
    \begin{tabular}{|p{0.1\textwidth}|p{0.5\textwidth}|}
\hline
\bf Attribute & \bf Complex Engineering Problems have characteristic P1 and some or all of P2 to P7\\
\hline
Depth of knowledge required & P1: Cannot be resolved without in-depth engineering knowledge at the level of one or more of K3, K4, K5, K6 or K8 which allows a fundamentals-based, first principles analytical approach\\
\hline 
Range of conflicting requirements & P2: Involve wide-ranging or conflicting technical, engineering and other issues \\
\hline 
Depth of analysis required
&P3: Have no obvious solution and require abstract thinking, originality in analysis to formulate suitable models\\
\hline
Familiarity of issues&P4: Involve infrequently encountered issues\\
\hline
Extent of applicable codes&P5: Are outside problems encompassed by standards and codes of practice for professional engineering\\
\hline
Extent of stakeholder involvement and conflicting requirements&P6: Involve diverse groups of stakeholders with widely varying needs\\
\hline
Interdependence&P7: Are high-level problems including many component parts or sub-problems\\
\hline
  \end{tabular}
    \label{tab:cep}
\end{table}

\begin{table}[!htb]
    \centering
    \caption{Range of Complex Engineering Activities}
   \begin{tabular}{|p{0.1\textwidth}|p{0.5\textwidth}|}
\hline
\bf Attribute&\bf Complex activities means (engineering) activities or projects that have some or all of the following characteristics.\\
\hline 
Range of resources&A1: Involve the use of diverse resources (and for this purpose resources include people, money, equipment, materials, information and technologies)\\
\hline 
Level of interaction&A2: Require resolution of significant problems arising from interactions between wide-ranging or conflicting technical, engineering or other issues\\
\hline 
Innovation&A3: Involve creative use of engineering principles and research- based knowledge in novel ways\\
\hline 
Consequences for society and the environment&A4: Have significant consequences in a range of contexts, characterized by difficulty of prediction and mitigation\\
\hline 
Familiarity&A5: Can extend beyond previous experiences by applying principles-based approaches\\
\hline 
  \end{tabular}
    \label{tab:cea}
\end{table}

Note that POc requires mapping with engineering design in a practice area and POe is mapped to engineering practice at the forefront of the discipline. Both of these two requirements if to be met must be accompanied with student experience related to engineering design and practice, preferably at the industrial standard. 

In addition to this, the definition of complex engineering problem as stated in Table~\ref{tab:cep}, requires P1 as mandatory and some or all of P2 to P7. Depth of Knowledge or P1 can not be resolved without in depth knowledge of one or more K3, K4, K5, K6 or K8. As POc and POe are already mapped to K5 and K6, if a curriculum ensures this, P1 could be satisfied in the design level of the curriculum. However, to ensure this the problems that are solved in the laboratory must require in-depth knowledge and also at least two other Ps. POj, which requires communication on complex engineering activities with engineering community and with society at large. 

With all this design in the curriculum, lets have a look at the assessment of this course. There had been a fixed assessment method and percentage of marks entitled to each course outcome. The assessment method with the percentage of marks is shown in Table~\ref{tab:assessment}.

\begin{table}[!htb]
\centering
    \caption{CO with Assessment Methods}
    \begin{tabular}{|p{0.1\textwidth}|p{0.28\textwidth}|p{0.1\textwidth}|}
    \hline \textbf{CO} & \textbf{Assessment Method} & \textbf{Marks(\%)} \\
    \hline - & Attendance & \hspace{0.6cm}10 \%\\
    \hline CO1 & Mid Exam (Hardware Setup)& \hspace{0.6cm}15 \%\\
    \hline CO1 & Final Quiz & \hspace{0.6cm}15 \%\\
    \hline CO2, CO3 & Project& \hspace{0.6cm}50 \%\\
    \hline CO3 & Class Performance &\hspace{0.6cm} 5 \%\\
    \hline CO3 & Lab Reports &\hspace{0.6cm} 5 \%\\
    \hline
    \end{tabular}
    \label{tab:assessment}
\end{table}

The emphasis had been put the most on project completion which was covered in CO2. Students were given a list of projects from which they selected a suitable one and sometimes offered new problem definitions and assessment of the project were based on demonstration of the final project. Usually it is a group project and towards the second half of the semester, student groups are supposed to show continuous updates on the project development culminated by a final demonstration in the last week of the semester. 

At the faculty level, a few drawbacks were identified for this curriculum and teaching-learning assessment. Firstly, the curriculum is a bit outdated as Arduino-based micro-controllers are not preferably used in the industry (violating K6, POe). Secondly, to conduct the experiments with Arduino, in-depth knowledge was no longer required, rather basic level of understanding was sufficient. Thirdly, the projects often lack depth of analysis, conflicting requirements, interdependence, etc., and thus not satisfying the required attributes of complex engineering problem solving. Lastly, the communication (i.e., report, presentation, demonstration, etc.) that are part of the assessment were not satisfying the requirements for the ranges of complex engineering activities. Based on these intuitive findings at the faculty level, as part of the  Continuous Quality Improvement (CQI) plan, a feedback mechanism from the stakeholders was initiated followed by curriculum revision. The steps are narrated in the following subsections.

\subsection{Feedback from the industry and academic stakeholders}
Before proposing our curriculum on Microprocessors and microcontrollers, we first decided to take feedback from the industry and academic experts who are working related to the application of different microprocessors, microcontrollers as well as IoT-related works. A total of 12 personnel from industry and academia were asked several questions regarding designing a new curriculum on Microprocessors and microcontrollers lab. Among the personnel, 75\% were from industry and 25\% from academia. The summary information of the stakeholders is shown in Table \ref{tab:feedback_ind}.

\begin{table}[!htb]
\centering
    \caption{Organizations that participated in the feedback process.}
    \begin{tabular}{|p{0.05\textwidth}|p{0.3\textwidth}|p{0.1\textwidth}|p{0.13\textwidth}|}
    \hline \textbf{SL No.} & \textbf{Domain/Field of Expertise} & \textbf{Company Type} & \textbf{Type of Customers}\\
    \hline 01. & Machine Learning, IoT & University & International\\
    \hline 02. & Data Science & University & International \\
    \hline 03. & UWP, Spring Boot & Industry & International\\
    \hline 04. & Natural Language Processing, Computer Vision & Industry & International\\
    \hline 05. & Software & Industry & International \\
    \hline 06. & Artificial Intelligence, Digital Logic Design, Microprocessor and Interfacing, Computer Organization and Architecture & University & Local \\
    \hline 07. & Machine learning  & Industry & Local \\
    \hline 08. & Software Development, Industrial IOT \& Software Collaboration & Industry & International \\
    \hline 09. & Android Development & Industry & International \\
    \hline 10. & Embedded System, C\# Dotnet, Web application & Industry & Local \\
    \hline 11. & IoT, Embedded Systems & Industry & International\\
    \hline 12. & Industrial Ecology & Industry & Local\\
    \hline
    \end{tabular}
    \label{tab:feedback_ind}
\end{table}

The stakeholders were provided the previous course outline and asked some specific questions regarding what types of content should be included in the updated curriculum. A sample question asked to the experts is shown in Table \ref{tab:industry_survey} along with the COs. 

\begin{table}[!htb]
\centering
    \caption{Sample Questions Asked to the Experts}
    \begin{tabular}{|p{0.05\textwidth}|p{0.25\textwidth}|p{0.3\textwidth}|}
    \hline \textbf{COs} & \textbf{Questions} & \textbf{Sample Answers}\\
    \hline CO1 & What are the types of microprocessors and microcontrollers used in your industry? & Raspberry Pi, AVR Microcontroller, ESP32 (RISC), ARM. Currently available all the MPU/MCU are used in the industry. \\
    \hline CO1 & What types of sensor data are monitored and analyzed? & Both analog and digital are used. Many sensors use standard protocols like I2C, SPI, etc. for communication\\
    \hline CO1 & What types of communication protocols are used in the industry? & Serial Communication, I2C(Inter-Integrated Circuit), SPI (Serial Peripheral Interface), RS485\\
    \hline CO2 & Are the problems that an engineer needs to solve always well defined? & Needs research to formulate the problem\\
    \hline CO3 & How do you train freshers to work with your system? & Train freshers by introducing them to the company, providing skill workshops, real-world projects, mentorship, and continuous learning. Encourage collaboration, feedback, and problem-solving.\\
    \hline
    \end{tabular}
    \label{tab:industry_survey}
\end{table}

The most important part of this lab is what type of microprocessors and microcontrollers should be taught in the course which is relevant to CO1. Therefore, we asked the stakeholders about the microprocessors and microcontrollers that are used in their industry/company as shown in Table~\ref{tab:industry_survey}. We also asked about their opinion on which microprocessors and microcontrollers should be included in the updated curriculum. A total of 84\% stakeholders mentioned RPi while 67\% mentioned AVR microcontroller and ARM. Among them, 59\% mentioned ESP32 as being used in IoT while all the stakeholders mentioned Arduino should be taught for a basic level of understanding. The results of this analysis are shown in \figurename\ref{fig:industry_microntler}. It is clear that RPi was the mostly suggested microcontroller by the industry and academic experts. For CO2, ``Design a basic embedded hardware group project'' was the intended output from this course. We intend to formulate this project's rubrics as an open-ended complex engineering problem that requires a literature review and other P's of a complex engineering problem. That is why we asked the relevant questions as shown in \ref{tab:industry_survey} for CO2. For CO3, ``Work in a team and communicate effectively'', we asked the questions specific to the freshers in the system as this basic course will build the base of the freshers who will later be involved in different hardware, IoT-based companies. 

\begin{figure}[h!]
    \centering
    \includegraphics[width=0.35\textwidth]{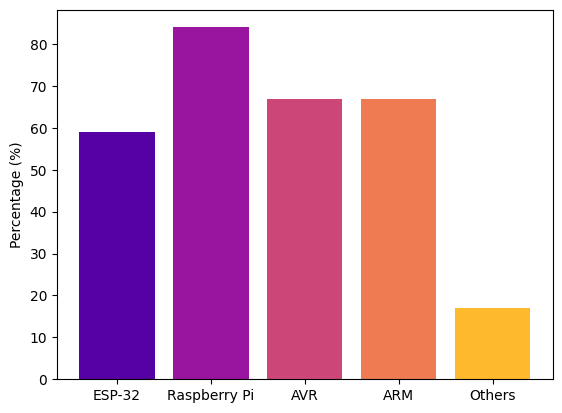}
    \caption{Different types of Microprocessors and Microcontrollers suggested by the Industry and Academic Experts. RPi is the most suggested one.}
%\caption{Example of a parametric plot}
    \label{fig:industry_microntler}
\end{figure}

According to other questions, the experts also suggested additional content that should be included in the proposed curriculum. These suggestions are shown in table \ref{table 4}.

\begin{table}[htbp]
\centering
    \caption{Relevant Content-wise suggestions from the experts to include in the curriculum}
    \begin{tabular}{|p{0.05\textwidth}|p{0.27\textwidth}|p{0.27\textwidth}|}
    \hline \textbf{SL No.} & \textbf{Specific Contents/Topics} & \textbf{Suggestions from the Experts}\\
    \hline 01. & Sensors that should be used & Different types Analog and Digital sensors\\
    \hline 02. & Types of communication protocol that should be taught & USerial Communication, I2C Communication, RS485\\
    \hline 03. & Machine Learning/Computer vision should be taught & Yes (according to all experts)\\
    \hline 04. & Growing need of IoT based projects & Should be included contents related to IoT\\
    \hline 05. & Open Ended Complex Engineering Problem & Should be included\\
    \hline
    \end{tabular}
    \label{table 4}
\end{table}

According to the feedback and suggestions from the experts, we decided to design a new curriculum, i.e. teaching-learning-assessment for this laboratory course while keeping in mind the limitations and prospects of the previous curriculum.

\subsection{Major Changes in the Curriculum}
In our previous curriculum, the AVR microcontroller and Arduino were taught. The previous curriculum included open-ended complex engineering projects carrying 50\% marks which should be completed by the end of the semester. Arduino is one of the basic microcontrollers that students can easily afford to solve complex projects due to its inexpensiveness. Thus, we decided to keep this microcontroller in the newly designed course as well. However, according to the experts, the growing need increases for the combination of machine learning and computer vision with microprocessors and microcontrollers. Therefore, we decided to use the RPi microprocessor while removing AVR from the updated course. RPi is suggested most by the experts, and very complex robotic projects can be accomplished using such a powerful microprocessor. We decided to design two experiments for this microprocessor. Also, the growing need for IoT in the industry is increasing and that is why we selected ESP32 while designing the new lab experiments.

\subsection{Development of New Curriculum and Analyzing the effect of this curriculum among students}
Analyzing the feedback from the stakeholders, we proposed a new curriculum. Then, we deployed this curriculum among the students in the Fall 2023 semester. To understand the impact of this curriculum we performed surveying. We also performed statistical analysis \cite{MainCronarticle, Ugoniarticle} on the students' course outcomes based on different evaluations such as Mid Exam, Final Quiz, open-ended projects, etc,.

\section{Proposed Curriculum} \label{s:propos-curcl}
The proposed curriculum is composed of four guided experiments and one open-ended project. However, the assessment method and the marks distribution have been kept the same as before (shown in Table \ref{tab:assessment}).

\subsection{Guided Experiments}
Guided Experiments are part of the close-ended approach in our proposed curriculum. These guided experiments are designed to teach students a basic understanding of microprocessors, microcontrollers, sensors, and how to integrate them. Additionally, the lab tasks and reports for these experiments are prepared to guide students in solving complex engineering problems. We refer to these experiments as 'guided' because they help students navigate complex engineering tasks and prepare them for open-ended lab projects.
In our proposed curriculum, there are four newly designed guided experiments based on Arduino, ESP module, IoT cloud, and RPi. Note that these experiments are assessed through Class Performance, Lab Reports, Mid Exam (Hardware Setup), and Final Quiz which carry a total of 40\% mark. The details of the experiments are given below.

\subsubsection{Experiment 1- an introduction to Arduino, Interfacing of Gas Sensor Using Arduino \& Showing the Sensor Data in OLED Display (weeks 1 and 2)} The objectives of this experiment are to learn about the Arduino boards and Arduino IDE, to demonstrate hardware and software interfacing using Arduino, to emulate Arduino projects in Proteus software to interface gas sensor with Arduino, and to show data in the OLED display.
    \begin{figure}[h!]
    \centering
    \includegraphics[width=0.35\textwidth]{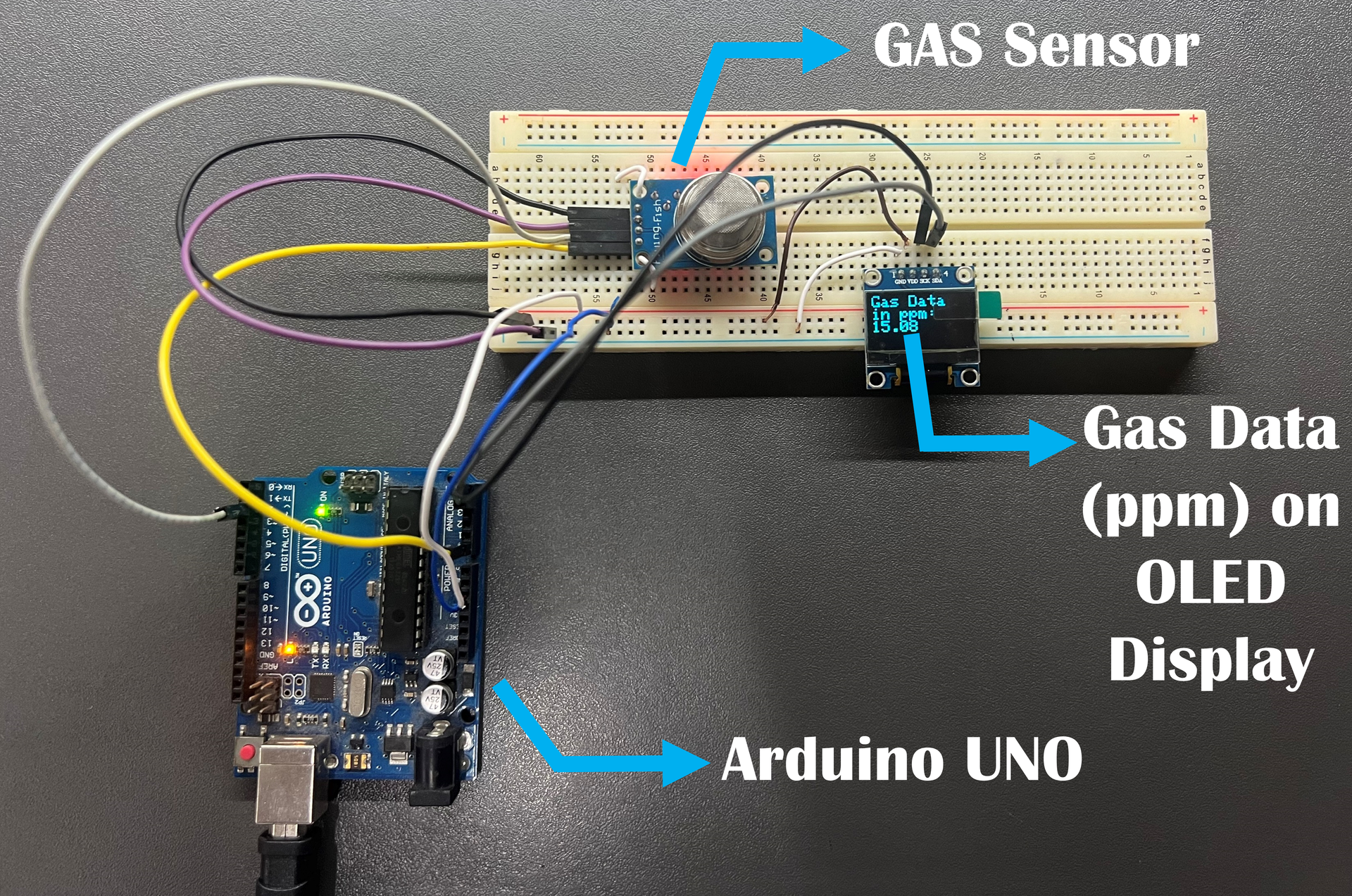}
    %\captionsetup{justification=centering}
    \caption{Hardware Connection of MQ2 Gas Sensor and OLED Display with Arduino Board \textcolor{black}{(Diagram of the experimental setup of Experiment 1 in the proposed curriculum)}.}
%\caption{Example of a parametric plot}
    \label{fig:arduino_sensor}
\end{figure}

Figure~\ref{fig:arduino_sensor} presents the hardware connection for this experiment. An Arduino Uno is connected \textcolor{black}{to} MQ2 gas sensor and the data is shown on the OLED display. Parts per million i.e., PPM, has been used as the unit of gas data. This experiment helps the students to handle real-life data and make them prepared to complete specific tasks in the industry. This experiment covers CO1 and CO2 in our curriculum. For further tasks such as lab reports, the students have to solve problems related to Light Dependent Resistors (LDR), SONAR sensor, and motorized exhaust fan-based gas sensor system which improve their critical thinking and programming skills. The students implemented both hardware and software parts in this experiment. The software implementation is based on Proteus simulation software.

\subsubsection{Experiment 2- Wi-Fi Communication and building IoT based systems using Arduino (week 3)} ESP-32 is used for gas sensor data monitoring and IoT data is transmitted to the Arduino cloud. The main advantage of this experiment is that the students can use Wi-Fi in their works without any peripheral module as ESP-32 has built-in Wi-Fi module. This experiment has great importance for building IoT-based systems.

    \begin{figure}[h!]
    \centering
    \includegraphics[width=0.35\textwidth]{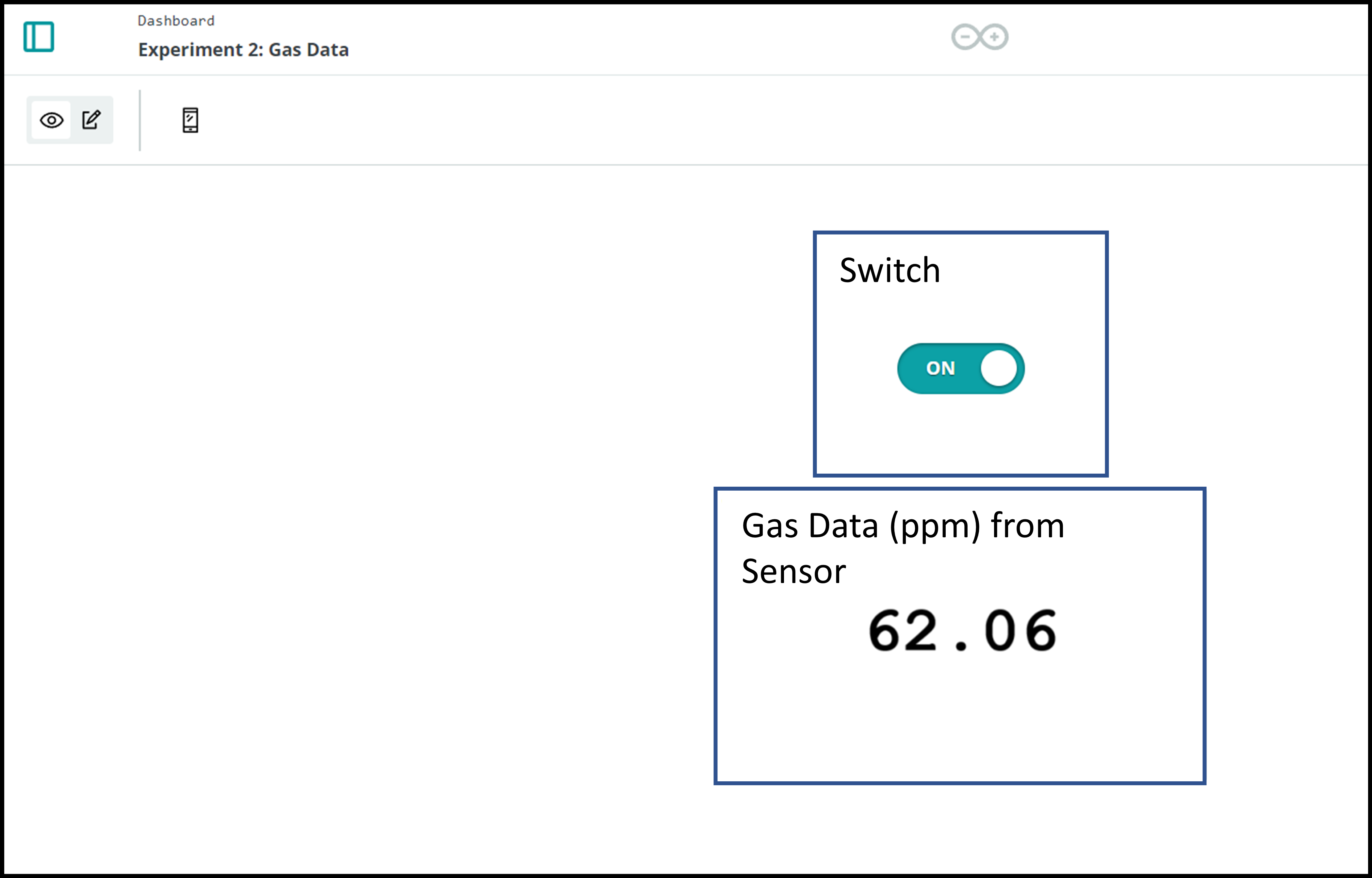}
   %\captionsetup{justification=centering}
    \caption{Arduino Cloud Dashboard Showing Data from Gas Sensor and Controlling a Switch \textcolor{black}{(Sample diagram from Experiment 2 in the proposed curriculum)}.}
%\caption{Example of a parametric plot}
    \label{fig:arduino_display}
\end{figure}

Figure~\ref{fig:arduino_display} presents an Arduino dashboard cloud that monitors and displays data from the MQ2 gas sensor. This data is sent directly from ESP-32 to the cloud and further tasks can be completed using this data. For additional tasks in this lab, the students have to extract or save the data of the gas sensor in a Google sheet or Excel file and activate a buzzer if the PPM value crosses a certain threshold (for instance, if methane's concentration goes above 1000 ppm). Additionally, they learn how to send text through the cloud and run a suitable system using the ESP32 local server. Further, the students have to build an IoT system where they can send moisture data from the ESP board to the cloud and control a water pump using those data. This task integrates the lesson of both experiments 1 and 2. After successfully conducting this experiment, students can effectively use cloud-based IoT dashboards to monitor and control real-time data. ESP 32 is one of the most common modules used in the IoT industry. This experiment covers CO1 and CO2 in our newly designed curriculum.

\subsubsection{Experiment 3- Introduction to Raspberry Pi (Gen 4 Model B/B+) (week 4)} The main objectives of this experiment are to introduce the RPi to the students, to set up the RPi operating system, and to run basic circuit codes using hardware on this system.

    \begin{figure}[h!]
    \centering
    \includegraphics[width=0.35\textwidth]{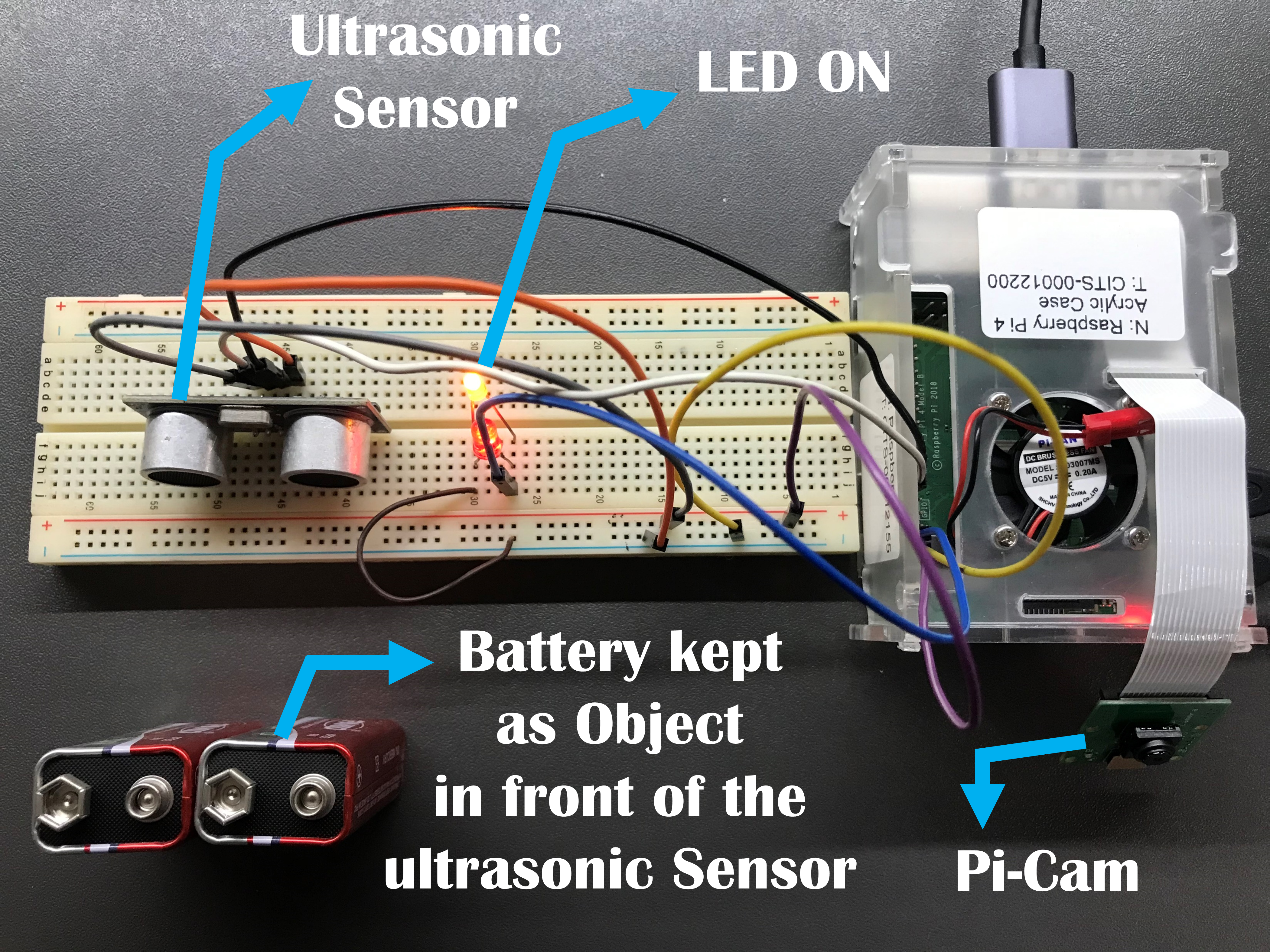}
   %\captionsetup{justification=centering}
    \caption{Distance Measurement Using Ultrasonic Sensor Interfacing With RPi \textcolor{black}{(Diagram of the experimental setup of Experiment 3 in the proposed curriculum)}.}
%\caption{Example of a parametric plot}
    \label{fig:pi_sensors}
\end{figure}
    
In \figurename~\ref{fig:pi_sensors}, a distance measurement system has been built using RPi and Ultrasonic Sensors. The ultrasonic sensor measures distance from a certain object, in this case, a battery. When the distance crosses a certain threshold, RPi activates the LED, alternatively, it is turned OFF. This experiment covers course outcomes of CO1 and CO2, of our newly designed curriculum. As tasks for lab reports, the students have to interface the DHT11 or DHT22 sensor with RPi to monitor humidity and temperature data and send the data to the I2C OLED display. After getting introduced to the basic RPi system, the students feel more confident trying advanced tasks such as the next experiment in our course.

\subsubsection{Experiment 4- Image and Video Processing with Raspberry Pi by interfacing Pi camera (weeks 5 and 6)} 
\begin{figure}[h!]
    \centering
    \includegraphics[width=0.35\textwidth]{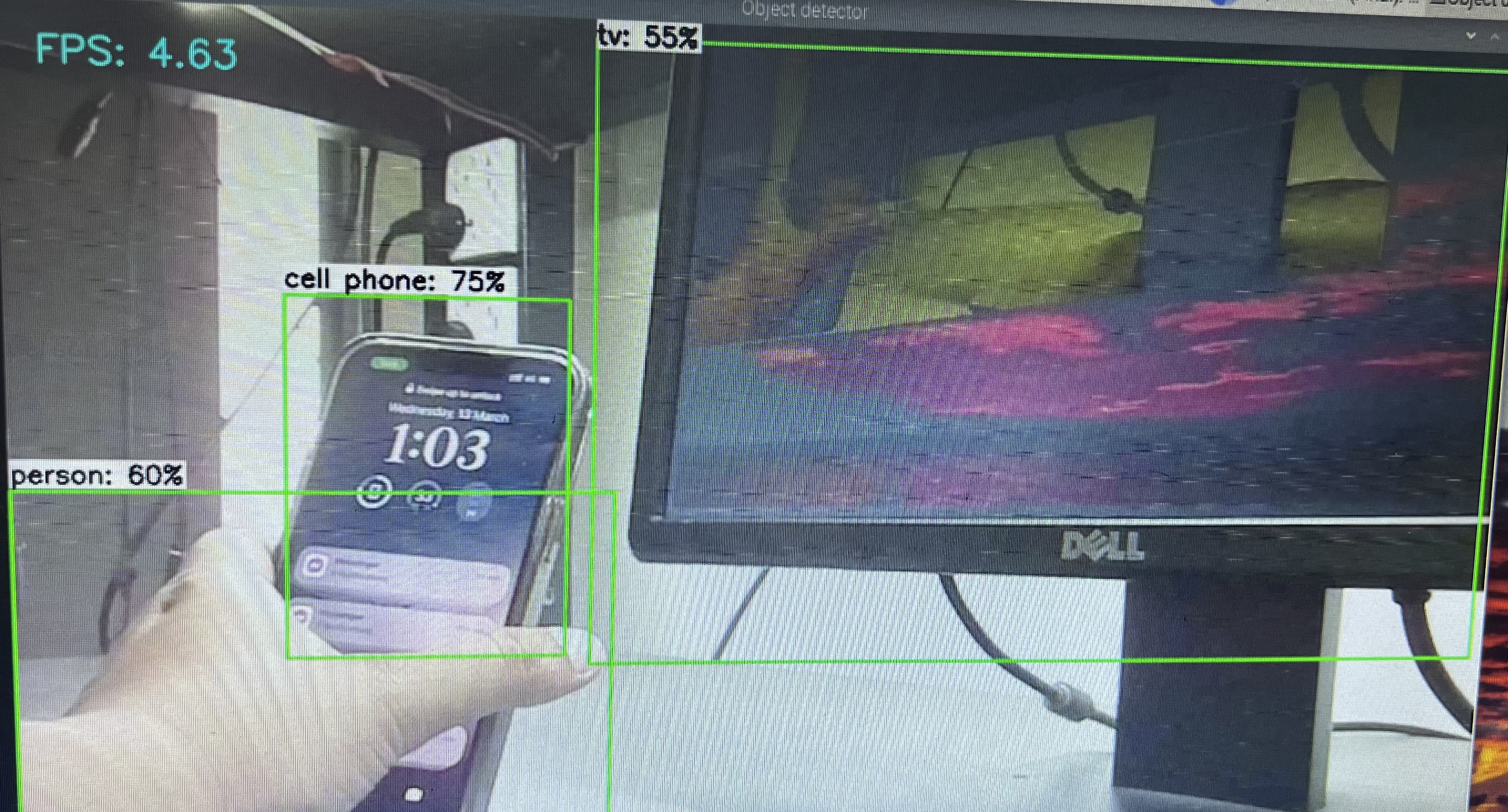}
   \captionsetup{justification=centering}
    \caption{Multiple Object Detection Using PiCam \textcolor{black}{(Sample diagram from Experiment 4 in the proposed curriculum)}.}
%\caption{Example of a parametric plot}
    \label{fig:pi_OD}
\end{figure}

In this experiment, the students learn how to use the camera module with RPi. They capture images, videos, and save those data in the system. They perform these activities using Python programming. Python's combination of simplicity, versatility, extensive libraries, and community support makes it a preferred choice for this experiment. Additionally, the RPi supports a default Python editor to communicate with the RPi I/Os. Further, they complete object detection using TensorFlow Lite and OpenCV which make them prepared to solve computer vision-related problems. Figure~\ref{fig:pi_OD} presents a part of the experiment where object detection has been demonstrated with an RPi Camera. The camera detects a cell phone, a monitor, and the hand of a person efficiently and simultaneously. The trained model effectively detects 80 common household objects. This experiment covers CO1 and CO2 in our curriculum. By completing the tasks in this experiment, the students gain practical experience on how to utilize the camera module in the RPi to handle images and video and process them using \textcolor{black}{a} machine learning toolkit. After learning the skills in this experiment, the students feel confident in solving industry-level problems of detection, classification, and identification of images and video frames.
    
\subsection{Open-ended Project Design}

In the newly designed curriculum, students in the  course embarked on a project-based learning experience (intended outcomes of CO2 and CO3), culminating in impact-full projects. During the second half of the semester, students have to complete an open-ended project based on their knowledge gained from the four guided experiments. The main objective of the projects is to solve a real-life problem effectively and efficiently. In the new curriculum, the students can include RPi and its communication protocols in their projects. However, use of RPi has not been declared compulsory due to its high cost. After shifting to the new curriculum, students have been vigilant in proposing more real-life projects. This is because the students now complete more advanced tasks in the guided experiments compared to the previous curriculum. Therefore, they feel more confident in solving even more complex engineering problems. A project rubric, accounting for 50\% of their final grade, guide their journey. Based on the feedback from the experts the design of the project has been updated. We present the updated project description as follows.

\subsubsection{Criteria of the project}
To ensure that each of the projects has the attributes required by complex engineering problem-solving standards and culminate the necessary POs, certain project criteria are set that the students have to follow to complete the project. The criteria have been set by the faculties in a way such that the students can tackle complex engineering problems by integrating both software and hardware knowledge learned from the experiments. These criteria will help the students to track their project updates keeping certain requirements in mind and encourage them to work as a team.
The details of the project criteria are given below: 

\begin{itemize}
    \item Arduino, AVR Microcontroller, NodeMcu or {RPi, ESP32} based embedded systems or IoT systems should be designed.
    \item Must use at least two sensors (analog or digital) to develop the system, a wireless data transferring system (depending on the problem’s specification), and at least one motor (e.g., DC motor or stepper motors).
    \item Show the necessary status/output relevant to the system in an OLED Display/website/ Mobile app etc.\ using wireless data transfer method, I2C data transfer method, etc. The overall system should be constructed such that it is a robust system.
    \item Must use different types of communication protocols (e.g., SPI, I2C, CAN Bus, or Serial Bus).
    \item Every member of the group must contribute to the hardware assembling step and in their respective codes to develop the overall system.
    \item The project must have at least four major features as every group consists of 4-5 students. For example, in a project aimed at building a Fire Fighting Robot using Computer Vision, the primary function is to detect fire within an area using an ESP32 camera. Other key features include navigating through the area after detection, stopping at a certain distance from the incident, measuring the intensity of the fire, and opening water valves to spray water and mitigate the situation. By specifying the main features in the project proposal, students can effectively work towards their targets on a week-by-week basis. These features are considered integral to their project.
    \item The project should be budget-constrained and cost-effective.
    \item {While Building the projects, students must estimate the number of I/O needed from the microcontroller. Based on this estimation students have to select the appropriate version of the microprocessor or microcontroller. Also, while building the system, same type of microcontroller or microprocessor can not be used multiple times if not required. For example, while Arduino Uno is selected at first. However, after implementing some features, it is observed that this microcontroller fails to provide with more I/Os to interface with other sensors and hardware. In such cases, students should first check if the hardware can be connected in a more organized way so that they do not need to buy another Arduino Uno considering the budget limit. However, if they still need another microcontroller, these two microcontrollers must be connected in a master-slave mode rather than using them as different entities.}
    \item {The overall project’s setup and design should be neat and clean. It should look as much industry grade as possible. Also, the controlling of the different features should be efficient.}
\end{itemize}

\subsubsection{Week-by-week task allocation} The project work will be mainly carried out from weeks 7 to 12.

\noindent \textit{Week 7: Preparation and Planning}

\begin{itemize}
    \item \textit{Literature Review:} Prior to project initiation, students conducted a thorough literature review. Their task was to identify gaps in existing research and propose novel solutions.
 \item \textit{Project Selection:} Faculty members assigned each group (4-5 members) a project with the potential for significant outcomes.
 \item \textit{Project Proposal Presentation:} In week seven, groups presented their project proposals, outlining their implementation plan.
 \item \textit{Interim Report:} Alongside the presentation, students submitted an interim report detailing their hardware and software choices. Faculty assessed their project objective analysis during this stage (10\% of total project marks).
\end{itemize}

\noindent \textit{Weeks 8-11: Development and Evaluation}

\begin{itemize}
    \item \textit{Weekly Updates and Evaluations (10\% each):} Throughout weeks 8-11, groups presented weekly updates showcasing their project's progress with each new feature implemented. Faculty members evaluated each group member's contribution to the relevant hardware or software aspects.
\end{itemize}

\noindent \textit{Week 12: Finalization and Presentation}

\begin{itemize}
\item \textit{Final Project Presentation:} During the final week, each group presented their completed project with a formal presentation, including (1) a GitHub repository link with project code and documentation, and (2) a 2-3 minute video showcasing all project features.
\item Final project report
\item \textit{Grading:} This final report, presentation, and demonstration are done for the rest of the assessment. The project completion is done in a form of project show conducted for this course in the last week of each semester where industry experts and faculty members participate as evaluators. This showcasing is an open event where fellow engineering students from the same discipline, other disciplines, and external guests are also invited for the demonstration. The students are engaged in communication with the audience (satisfying the culmination requirements for POj). 
\end{itemize}

This project-based learning approach fosters valuable skills, including literature review, critical analysis, project management, and interpersonal communication. \textcolor{black}{The reasonability of different projects are evaluated following the criteria above. In different weeks of the project, the faculties ensure that the students are working properly and meeting the requirements. These weekly evaluations following complex engineering problem formulation are then reflected on the final evaluations.}
The emphasis on weekly updates ensured continuous progress and feedback from faculty members, leading to impact-full project outcomes thus culminating all the POs.

\subsection{Satisfying the Requirements of a Complex Engineering Problem}

In the project report, they wrote the literature review, explained their proposed method, wrote about the implementation of hardware and software, analyzed their results, and discussed how their project performed in real-life scenarios. They wrote the report in LaTeX using the IEEE template. In the discussion section of the report, they had to answer complex engineering problem mapping questions where they would explain if they gained a depth of knowledge (\textbf{P1}), if they faced conflicting requirements (\textbf{P2}), if there was any better alternate solution or design other than the one they have implemented (\textbf{P3}), the impact of their solution (\textbf{P4}), if they maintained ethical code of conducts while implementing their project (\textbf{P5}), if the stakeholders' requirements were incorporated in their solution (\textbf{P6}), if the modules in their system had interdependence or not (\textbf{P7}). Finally, they had to put tic marks in the following Table~\ref{tab:cep_report} where it is applicable and provide rationale and evidence.

\begin{table}[!htb]
    \begin{center}
    \caption{Complex Engineering Problem Mapping. %Tic P1 and some or more from P2-P7 as applicable. P1 - Depth of Knowledge, P2 - Conflicting Requirements, P3 - Depth of Analysis, P4 - Familiarity of Issues, P5 - Extent of Applicable Codes, P6 - Extent of Stakeholders, P7 - Interdependence.
    }
    \label{tab:cep_report}
    \begin{tabular}{|c|c|c|c|c|c|c|}
\hline  
P1&P2&P3&P4&P5&P6&P7\\
\hline 
$\checkmark$&$\checkmark$&&&&&$\checkmark$\\
\hline 
\end{tabular}
\end{center}
    
\end{table}

As the students are required to satisfy the criteria of the project before the topic is selected, in most cases, P2, and P7 are satisfied as they are required to work under constraints or conflicting requirements and multiple interdependent components must be present in the design. In addition to these often students select problems with application in the area of agriculture, health, manufacturing, etc which allows the culmination of POc as well as a few other Ps (P4, P5, P6). All these are also included in the report, presentation, and demonstration during the project show which ensures the related complex engineering activities are ensured through effective communication.

The final week contains 30 marks which is 60\% of the total project marks. They are evaluated through their presentation, question and answer, project report, overall project status, and project show performance. Overall project status is considered satisfactory if all the features are implemented neatly (engineering practice). However, the projects in the previous curriculum were not up to industry standards according to industry experts as the students mainly focused on Arduino-related projects. Many complex engineering problems could not be solved using only Arduino. Consequently, introducing advanced microprocessor-based systems such as RPi became very essential as per the survey we have conducted in the industry and academia. By introducing RPi, the students can approach industry-level problems. Next, we will see the survey we have conducted among the industry and academic experts about what types of experiments we should conduct, and according to that, we will propose a new updated course curriculum.

\subsection{Feedback on the Proposed Curriculum}

\begin{figure}[h!]
    \centering
   \includegraphics[width=0.50\textwidth]{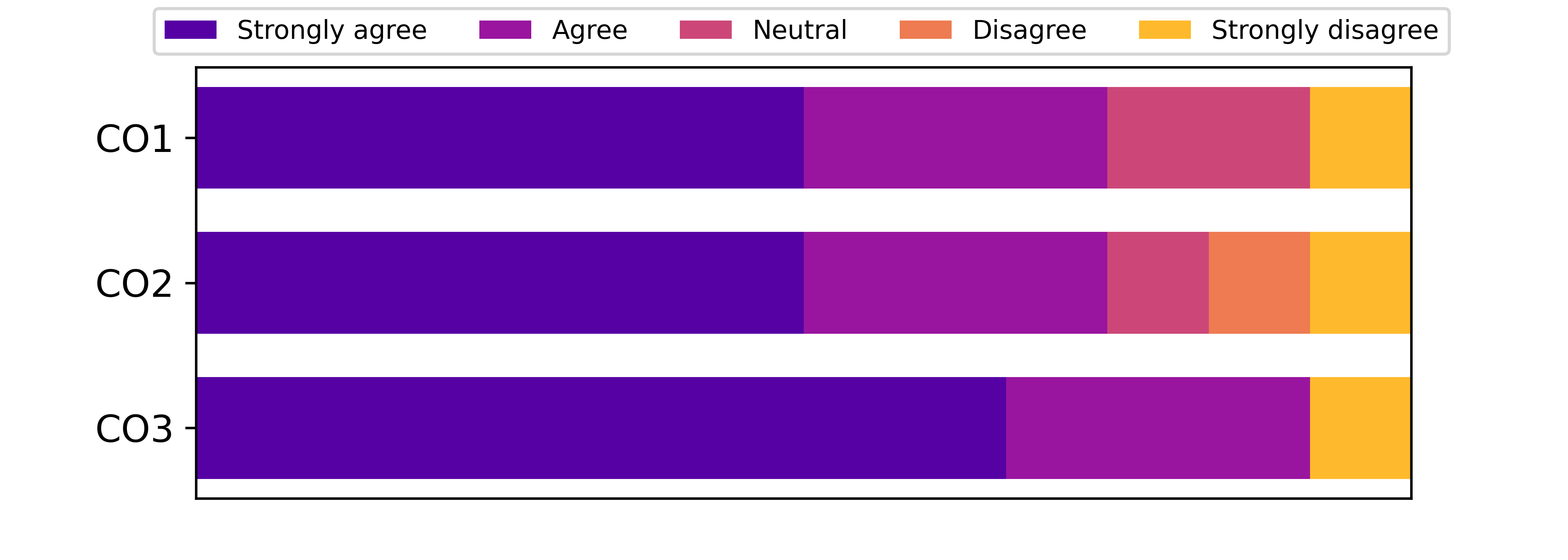}
    \caption{Feedback from the experts on the proposed curriculum based on the COs.}
%\caption{Example of a parametric plot}
    \label{Figure 2}
\end{figure}

After designing the course according to the industry needs and experts' feedback, we sent the detailed course outline to the same experts and asked their opinion on the overall design of our experiment based on the COs of this course. They were asked whether our newly designed course can justify the course outcomes. Their quantitative opinion on the newly designed course is shown in Figure~\ref{Figure 2}. We can see that the experts broadly agree with our proposed design based on the COs. They also mentioned that more information needs to be gathered by deploying this course to the students to justify CO2. They have expressed that we can deploy this course to the students and can make some changes in CO2 in the future after getting some quantitative feedback from the students. 

\section{Results and Analysis} \label{s:result-analys}
The newly proposed curriculum was introduced in the Fall 2023 semester. \textcolor{black}{In this course, there are roughly 30 to 35 students in each section. We propose that our suggested curriculum can work well in classes with 30 to 35 students as well as those with less than 30. However, a variety of elements, like the number of faculty members and enough infrastructure, are crucial when it comes to class size.}The students were asked several questions for a survey at the end of the semester. They were also evaluated based on their class performance, mid-exam and Final Quiz, Laboratory reports, and finally on a complex open-ended problem which they proposed to do at the beginning of the semester. In this section, we discuss the survey and different statistical results on the evaluation marks to understand the impact of the proposed curriculum.

\subsection{Student Survey Results}
The students were asked a total of 13 questions relevant to the impact analysis of the proposed experiments. In total, 33 students participated in this survey. They gave a rating on these questions (5 indicating strongly agree and 1 indicating strongly disagree). The questions asked in the survey are given in Table \ref{tab:student_survey} along with their average marking by the students. 

\begin{table}[!htb]
\centering
    \caption{Survey questions asked to the students for the proposed curriculum and their average rating}
    \begin{tabular}{|p{0.1\textwidth}|p{0.38\textwidth}|p{0.09\textwidth}|}
    \hline \textbf{Q. No.} & \textbf{Questions} & \textbf{Average Number (1-5)}\\
    \hline 01. & Do you think that CO1 is achieved in this Lab? & 4.54\\
    \hline 02. & Do you think that CO2 is achieved in this Lab? & 4.58\\
    \hline 03. & Do you think that CO3 is achieved in this Lab? & 4.48\\
    \hline 04. & Do you think that the course instructor delivered the lectures based on the COs? & 4.64\\
    \hline 05. & Is the course material and assessment aligned with the course COs? & 4.67\\
    \hline 06. & Do you think that all of the COs were covered in the Course? & 4.52\\
    \hline 07. & Were real-life scenarios and industry-level applications using microprocessors and microcontrollers discussed during delivering the course lectures? & 4.42\\
    \hline 08. & How exciting was your experience while working with I/O devices such as sensors, LEDs, or motors while building a microcontroller-based project? & 4.52\\
    \hline 09. & How much did you enjoy simulating a specific electronic hardware circuit in Proteus? & 4.52\\
    \hline 10. & How exciting was the application of microcontrollers such as ESP32 in IoT projects, such as home automation or wearable gadgets? & 4.52\\
    \hline 11. & How much did you enjoy the application of microprocessors such as RPi in projects like  Machine Learning project, image processing, etc? & 4.48\\
    \hline 12. & During the course, how important was effective communication among team members to achieve project success? & 4.48\\
    \hline 13. & Rate your thoughts on how applicable the knowledge and skills gained in this course are to solve real-world problems & 4.30\\
    \hline
    \end{tabular}
    \label{tab:student_survey}
\end{table}

\begin{figure}[h!]
    \centering
    \includegraphics[width=0.50\textwidth]{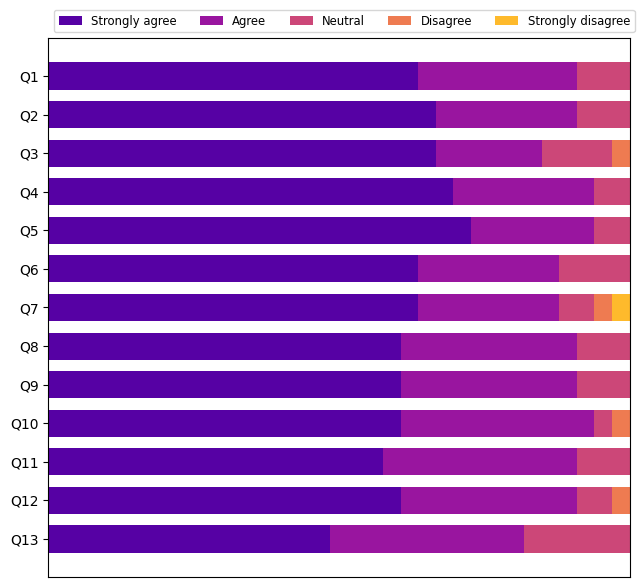}
    \caption{Spectrum of responses in the student survey for the proposed curriculum.}
%\caption{Example of a parametric plot}
    \label{fig:13q}
\end{figure}

As the course outcomes (COs) are one of the important aspects of an engineering course, we intended to ask several questions related to the COs to the students in the survey. They were briefly discussed about the COs at the start of the semester as well as before the survey. In Table \ref{tab:student_survey}, we can see that the COs achieved very good average ratings according to the students. We also included some questions related to students' perception of the courses and which parts they were enjoying the most. The completion of an open-ended project is one of the main objectives of this course. As a result, we asked questions no. 12 and 13. The average rating of all the questions is above 4, indicating that the students enjoyed the lab experiments as well as understanding the impact of the proposed curriculum in building open-ended projects. Now, We also indicated that rating 5 in questions means strongly agree and 1 means strongly disagree with the notion of the question. We plotted a 2D bar graph (\figurename\ref{fig:13q}) showing the spectrum of responses to understand the in-depth analysis of the survey besides average ratings.

We note from \figurename\ref{fig:13q} that most of the students agreed or strongly agreed with the statements. Question no. 13 received less ``strongly agree'' compared to the other questions. This may be due to the less knowledge of the students about the trends in hardware and software-related works in real life and industry. Additionally, we assessed the internal consistency of this student survey using Cronbach's $\alpha$ reliability measure. Cronbach’s $\alpha$ assesses the internal consistency of a set of scale or test items \cite{MainCronarticle,Cronarticle}. If the value of $\alpha$ exceeds or equals a threshold of 0.7, the internal consistency is considered satisfactory. Higher values of $\alpha$ indicate increasing levels of internal consistency \cite{Cronarticle}. For a total of 13 survey questions, 26 students answered which is an acceptable sample size for Cronbach's $\alpha$ test. The value of $\alpha$ was found to be 0.93 for 99\% confidence level which indicated the survey questions' internal consistency as "Excellent".

\subsection{Statistical Analysis}
The evaluation for the students in this course was always computed according to the three COs where CO1 contains 30 marks (Mid Term and Final Exam), CO1 and CO2 contain 50 marks (Open-ended complex engineering project), CO3 contains 10 marks (Class performance, group activities, and Lab reports) and finally 10 marks for the class attendance. We have taken the COs-based marks of students completing this lab in Spring 23, Summer 23, Fall 23 and \textcolor{black}{Spring 24} where the proposed curriculum was introduced  \textcolor{black}{in Fall 23}. Then, we computed the average percentage of marks obtained in these COs for different semesters and analyzed these data for $t$-tests to understand the significance of the newly proposed curriculum. The $t$-tests is a statistical method used to determine if there is a significant difference between the means of two groups while considering the null hypothesis that there is no difference between the means of the paired groups \cite{Ugoniarticle}. After calculating the $t$-value of each group of data, we can calculate the respective probabilities, called the $p$-value from the $t$- distribution. It is common practice to reject the hypothesis when the $p$-value is less than 0.05 and to retain it when the $p$-value is greater than 0.05 \cite{Ugoniarticle}. While conducting literature reviews, we observed that many articles have utilized $t$-tests to assess the significance of their approach while designing a new curriculum \cite{21, chiu2021creation}. In our case, these two groups can be referred to as the CO-based marks achieved by each student in the trimester before introducing the proposed curriculum and after introducing the proposed curriculum. We take the hypothesis as there is no significant difference in each COs before and after introducing the proposed curriculum. We can reject this hypothesis and conclude significant changes are observed if the $p$-value becomes less than 0.05. 

\begin{figure}[h!]
    \centering
    \includegraphics[width=0.35\textwidth]{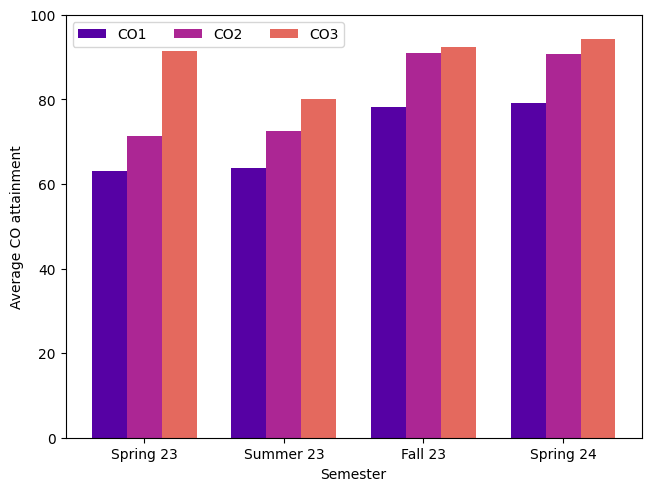}
    \caption{Impact Analysis of the new course outline from the average COs achieved. \textcolor{black}{We can observe that achievment in COs increased in Fall 23 after introducing the new curriculum}.}
%\caption{Example of a parametric plot}
    \label{fig:co_attainment}
\end{figure}

The average percentage of the COs achieved is shown in Table~\ref{tab:CO_attainment} and \figurename~\ref{fig:co_attainment}. We can see previously that CO1 and CO2 were not achieved significantly. However, after introducing our newly proposed curriculum in the Fall 2023 semester, CO1 and CO2 were achieved in higher percentages, approximately 15\% in CO1 and 18\% in CO2. As a result, the overall average CO increased approximately by 10\%-12\%. \textcolor{black}{Similarly, we can observe higher increase in COs in Spring 24 as demonstrated in Table \ref{tab:CO_attainment} and Figure~\ref{fig:co_attainment}}. The most significant improvement is in CO2 which is the open-ended complex engineering group project. This indicates that our newly proposed lab experiments have enabled students to perform complex engineering projects more competently due to the addition of the high-level course outlines and tasks taught in the lab. As a result, they are achieving higher marks in this category.

\begin{table}[!htb]
\centering
    \caption{COs-based achievement in consecutive semesters (Proposed curriculum introduced in Fall 2023 semester)}
    \begin{tabular}{|p{0.16\textwidth}|p{0.12\textwidth}|p{0.12\textwidth}|p{0.12\textwidth}|p{0.12\textwidth}|}
    \hline \textbf{Course Outcomes} & \textbf{Outcome Achieved (\%) in Spring 23} & \textbf{Outcome Achieved (\%) in Summer 23} & \textbf{Outcome Achieved (\%) in Fall 23} & \textcolor{black}{\textbf{Outcome Achieved (\%) in Spring 24}}\\
    \hline CO1 & 63.025 & 63.84 & 78.205 & \textcolor{black}{79.25}\\
    \hline CO2 & 71.24 & 72.42 & 90.92 & \textcolor{black}{90.795}\\
    \hline CO3 & 91.47 & 80 & 92.308 & \textcolor{black}{94.215}\\
    \hline \textbf{Average} & \textbf{75.24} & \textbf{72.089} & \textbf{87.15} & \textcolor{black}{\textbf{88.09}}\\
    \hline \textbf{Students evaluated} & \textbf{34} & \textbf{33} & \textbf{26} & \textcolor{black}{\textbf{33}}\\
    \hline
    \end{tabular}
    \label{tab:CO_attainment}
\end{table}

We have also performed paired t-test analysis on the CO distribution between the Spring 2023 and Fall 2023 semesters, shown in Table~\ref{tab:ttest}. We see a significant change in CO1 and CO2 ($t$=-3.976, $p$$<$0.05 and $t$=-6.529, $p$$<$0.05). The test score indicates that students have achieved more enhanced knowledge and application of the advanced applications of the microprocessors and microcontrollers from the new course outline which helped them to perform well in the MID and Final exam. They have also performed well in designing, and researching the open-ended complex engineering problem.

\begin{table}[!htb]
\centering
    \caption{t-test result between Spring 23 and Fall 23 semesters}
    \begin{tabular}{|l|l|l|l|l|l|l|}
    \hline \multirow{2}{*}{\textbf{COs}} & \multicolumn{2}{|l|}{\textbf{Spring 23}} & \multicolumn{2}{l|}{\textbf{Fall 23}} & \textbf{t-value} & \textbf{p-value} \\
    \cline{2-5} & \multicolumn{1}{c|}{\textbf{M}} & \multicolumn{1}{c|}{\textbf{SD}} & \multicolumn{1}{c|}{\textbf{M}} & \multicolumn{1}{c|}{\textbf{SD}} & &\\
    \hline \textbf{CO1} & 22.06 & 5.56 & 23.46 & 3.86 & $-3.976^{\star *}$ & 0.0001964 \\
    \hline \textbf{CO2} & 35.62 & 14.43 & 45.46 & 3.71 & $-6.529^{\star *}$ & $3.27 \mathrm{e}-08$ \\
    \hline \textbf{CO3} & 9.15 & 1.52 & 9.23 & 1.56 & 0.4093 & 0.6847 \\
    \hline \multicolumn{7}{|l|}{ M: Mean, SD: Standard Deviation, ${ }^{* *} \mathrm{p}<0.05$} \\
    \hline
    \end{tabular}
\label{tab:ttest}
\end{table}

A similar analysis was performed on the CO distribution between the Summer 23 and Fall 23 semesters, shown in Table~\ref{tab:ttest2}. We see a significant change in all the CO: CO1, CO2, and CO3 ($t$=-3.074, $p<$0.05; $t$=-4.434, $p<$0.05 and $t$=-2.88, $p<$0.05). In both cases, as shown in Tables~\ref{tab:ttest} and~\ref{tab:ttest2}, we find CO1 and CO2 as significant where they carry a total percentage of 80\% marks of this course. This shows the impact of the newly designed course integrating with machine learning and computer vision concepts.

\begin{table}[htbp]
\centering
    \caption{t-test result between Summer 23 and Fall 23 semesters}
    \begin{tabular}{|l|l|l|l|l|l|l|}
    \hline \multirow{2}{*}{\textbf{COs}} & \multicolumn{2}{|l|}{\textbf{Summer 23}} & \multicolumn{2}{l|}{\textbf{Fall 23}} & \textbf{t-value} & \textbf{p-value} \\
    \cline{2-5} & \multicolumn{1}{c|}{\textbf{M}} & \multicolumn{1}{c|}{\textbf{SD}} & \multicolumn{1}{c|}{\textbf{M}} & \multicolumn{1}{c|}{\textbf{SD}} & &\\
    \hline \textbf{CO1} & 19.15 & 6.27 & 23.46 & 3.86 & $-3.074^{\star *}$ & 0.00324 \\
    \hline \textbf{CO2} & 36.21 & 11.22 & 45.46 & 3.71 & $-4.434^{\star *}$ & $6.821 \mathrm{e}-05$ \\
    \hline \textbf{CO3} & 8 & 2.012 & 9.23 & 1.56 & $-2.88 {\star *}$ & 0.006289 \\
    \hline \multicolumn{7}{|l|}{ M: Mean, SD: Standard Deviation, ${ }^{* *} \mathrm{p}<0.05$} \\
    \hline
    \end{tabular}
\label{tab:ttest2}
\end{table}

\subsection{Qualitative Analysis of the Open-Ended Project Design}
One of the main approaches in our lab design is to motivate students to formulate complex engineering problems that can be solved in various ways using embedded hardware systems. We proposed this new curriculum by integrating modern microprocessors, microcontrollers with different communication protocols, sensors, and lastly computer vision so that the students can fulfill their main objectives of formulating such complex engineering projects. According to the final project rubrics, they are first required to conduct thorough literature reviews on the engineering problems they intend to solve. After that, the gap analysis should help them formulate the problem in a research-oriented way as instructed by the industry and academic experts (CO2). Previously, we only taught basic microcontrollers like Arduino, AVR but now due to the addition of ESP32 and RPi, the research area on complex engineering-related problems broadens as ESP32, and RPi can solve many advanced-level problems. That is why the average of CO2 achieved is very high and the t-test shows significance in CO2. We further explain the impact of our proposed curriculum on CO2 by showing the top projects performed by the students before and after introducing the newly proposed curriculum as shown in Table~\ref{tab:projects}.

\begin{table}[htbp]
\centering
    \caption{Sample Open-ended projects proposed and solved by the students before and after introducing the newly proposed curriculum}
    \begin{tabular}{|p{0.2\textwidth}|p{0.45\textwidth}|}
    \hline \textbf{Projects before the proposed curriculum} &  i) Automatic book management system/Book Vending machine using Arduino that can provide registered people with books using the app.
    
    ii) Smart Office system prototype where only employees can enter with ID card (RFID Card). The light and fans can be controlled automatically or manually using a controller interfaced with Arduino. The employees can take coffee by using their RFID cards.

    iii) IoT-based hardware management system using Arduino. This is one kind of vending machine for hardware systems. Different hardware components are stored here and people can take hardware components from this vending machine using an app or ID card. Only registered people can take the hardware components from the System/Box. People have to return this hardware component within a certain day which is controlled by an automated process.

    \\
    \hline \textbf{Projects after the proposed curriculum} & 
    i) Underwater Rover using ESP32 that has 4 degrees of freedom and can stream live videos using ESP32-cam as well as perform object detection.
    
    ii) Fire fighting robot using computer vision: The robot car can detect fire within a room/small surroundings using an ESP32 camera. After detecting the location of the camera, it can go to a particular location and stop at a certain distance after measuring the intensity of the fire. It can spray water to stop small fire hazards.

    iii) Computer vision-based electronic component management system where the electric components can be detected using a real-time camera. Registered people can take and return the components and the entire process can be verified using video analysis. The students completed this project for a few electric components such as motors, multi-meters, sensors, and batteries. The issuing of the components and returning them is completely automated.
\\
    \hline
    \end{tabular}
    \label{tab:projects}
\end{table}

We can see from the \ref{tab:projects} that the quality of the complex engineering projects increased significantly due to the integration of microprocessors with IoT as well as Computer vision to solve more real-life problems. We observe that projects completed before the proposed curriculum were Arduino-based which doesn't require in-depth knowledge anymore compared to the other microprocessors and microcontrollers, hence not satisfying the P1 characteristic of the complex engineering problems. On the other hand, previous projects' features were not accounted for to fulfill the other range of activities from P2 to P7. For example, project example before the proposed curriculum such as, `Automatic book management system', the student implemented an app via registered people can take books from a Box. This project did not address conflicting requirements (P2) or undergo a depth of analysis (P3) since it was a simple hardware project with obvious solutions. Furthermore, it lacked engagement with other complex engineering activities. Conversely, after the implementation of the proposed curriculum, students were provided with a set of rubrics outlining project criteria. They were required to propose projects and explain in their project reports which range of activities their projects fulfilled. For example, the project 'Underwater Rover' demanded in-depth analysis as students had to consider factors such as the rover's shape and size for successful underwater operation. Additionally, communication underwater posed challenges with no readily apparent solutions, especially considering budget constraints. Thus, this project addressed both P2 and P3. The students also explained these activities being achieved in the project report.  
\textcolor{black}{Our updated curriculum's guided experiments are made to be compatible with the open-ended stage. These guided experiments give students the knowledge they need to finish the tasks and get past any complexity that may come up during their projects by teaching them about different kinds of sensors, how their data is transferred using various communication protocols, and how various microprocessors and microcontrollers can be used for high-level integration of embedded systems.}In this way, all the projects after the proposed curriculum addressed the complex engineering activities, enabling students to tackle more intricate, industry-level problems.
\section{Conclusion} \label{s:con}
In this work, we adopted an integrated approach to teaching microprocessors and microcontrollers to undergraduate computer science and engineering students. We developed a series of guided lab experiments, ranging from basic to advanced levels, utilizing platforms such as Arduino, ESP32, and Raspberry Pi (RPi). These experiments were designed based on feedback from both industry professionals and academic experts and incorporated a variety of sensors (from simple to complex), IoT concepts, and elements of machine learning. This hands-on approach equips students with the skills needed to tackle open-ended, real-world projects that address complex engineering challenges.

Following the deployment of this course, student feedback was overwhelmingly positive, with notable improvements observed in key course outcomes (CO1 and CO2). Statistical analyses, including paired t-tests, confirmed these significant advancements. Looking ahead, we plan to extend this methodology to other senior-level CSE courses. \textcolor{black}{Higher level non-laboratory courses such as 
 {Green Computing, Digital Image Processing, Introduction to Bioinformatics, Machine Learning, Artificial Intelligence} can adapt this open-ended method of learning which are closely aligned with industry applications. Such a learning method will ensure that students can apply artificial intelligence techniques to tackle real-world challenges such as computer vision, healthcare, and automation. Since industry trends change quickly, this open-ended learning approach will ensure that students stay updated with industry norms.
} We believe the insights gained from this course can also be applied effectively to other engineering disciplines, broadening its impact across related fields.

\end{document}